# AI-driven Large-scale Electron Microscopy enables Whole-tissue Subcellular Digitization

Li Xiao[1]†*, Liqing Liu[2,3]†*, Hongjun Wu[1]†, Jiayi Zhong[4], Xixia Li[2], Yan Zhang[5], Junjie Hu[4,5]*, Sun Fei[2,5,6]*, Ge Yang[7]*, Tao Xu[5,8,9]*

[1]Beijing University of Posts and Telecommunications; Beijing 100876, China

[2]Center for Biological Imaging, Institute of Biophysics, Chinese Academy of Sciences, Beijing 100101, China

[3]Interdisciplinary Center for Biointelligence, Institute of Biophysics, Chinese Academy of Sciences; Beijing 101407, China

[4]School of Life Sciences, University of Chinese Academy of Sciences, Beijing 100049, China

[5]State Key Laboratory of Biomacromolecules, Institute of Biophysics, Chinese Academy of Sciences, Beijing 100101, China

[6]China-New Zealand Joint Laboratory on Biomedicine and Health, Guangzhou Institutes of Biomedicine and Health, Chinese Academy of Sciences, Guangzhou 510530, Guangdong, China

[7]State Key Laboratory of Multimodal Artificial Intelligence Systems, Institute of Automation, Chinese Academy of Sciences, Beijing 100190, China

[8]School of Biomedical Engineering, Guangzhou Medical University, Guangzhou 511436, Guangdong, China

[9]Guangzhou National Laboratory, Guangzhou, 510005, China

†These authors contributed equally to this work.

*Corresponding author.

Email: andrewxiao@bupt.edu.cn;liuliqing@ibp.ac.cn;huj@ibp.ac.cn;feisun@ibp.ac.cn； yangge@ucas.edu.cn;xutao@ibp.ac.cn

**Abstract:** The distribution and interactions of cellular organelles play a critical role in mediating cellular physiology and pathology. Large-scale electron microscopy enables visualization of organelle distribution and interactions at the tissue level with nanometer resolution, but robust and efficient computational analysis tools are lacking. Here, we present a deep learning tool for universal large-scale 2D/3D electron microscopy analysis, DeepOrganelle. This new tool enables high-throughput, cell-resolved spatiotemporal mapping and digitization of organelle distribution and interactions. When applied to spermatogenesis across 12 stages and 22 differentiation status of the germ cells, DeepOrganelle uncovered previously unrecognized, stage-dependent dynamics of mitochondria-endoplasmic reticulum contact sites within one subphase of prophase I during meiosis. It also revealed coordinated organelle redistribution in Sertoli cells towards the blood–testis barrier, digitizing the remodeling dynamics of the tissue. This study demonstrates that DeepOrganelle provides a powerful framework that captures subcellular dynamics at the whole-tissue level.

## Introduction

Eukaryotic cells are compartmentalized by intracellular membranes into organelles with distinct physiological functions. The subcellular architecture structurally defines the functional state of the cell. Digitizing cells by mapping their intricate organellar networks onto their physiological and pathological contexts has the potential to contribute substantially to the development of digital cell models, also referred to as virtual cell models, and to the mechanistic understanding of cell atlases. This digitization approach enables the identification of cell type-specific subcellular organization signatures and patterns, offering cross-scale insights into fundamental biological processes and disease mechanisms.

However, current efforts to construct "digital cells" are driven predominantly by single-cell or spatial multi-omics technologies. Recently, spatially resolved multi-omics atlases have made substantial advances across diverse biological systems [1-7], transforming our understanding of tissue organization—from organ-scale architecture down to cell type-specific molecular states. Yet, these approaches generally lack resolution at the subcellular level. Organelles and their dynamic interactions, mediated by structures such as membrane contact sites (MCSs), are basic constituents of a truly complete digital cell. To achieve cross-scale cellular digitization, it is necessary to quantitatively map organelle architectures and interactions inside cells within their native tissue contexts. However, digital cells [8-10] offer unprecedented tools to simulate and predict cellular responses to genetic or pharmacological perturbations, establishing a new paradigm for understanding disease mechanisms, predicting drug treatment effects, and developing patient digital twins. Nevertheless, because of their reliance on single-cell omics [11-13], current virtual cells suffer from limited spatial context, static characterizations, and a lack of physical parameters. This may hinder the creation of predictive, high-fidelity models. Volume electron microscopy (vEM) delivers nanoscale 3D structural maps that represent native spatial architecture and quantify physical features, such as organelle morphology and organelle interactions. By combining omics with vEM data, digital cell models can gain spatial information, subcellular representation, and physical parameters, evolving from static single-cell profiles into dynamic predictive and subcellular-resolution systems.

Recent advances in vEM [14,15] motivated by connectomics studies [16-18] and propelled by breakthroughs in high-speed imaging technologies, such as multi-beam scanning electron microscopy (EM) [19,20], have enabled nanoscale-resolution imaging at millimeter scale across tissues, entire organs, and even whole small model organisms. These capabilities simultaneously capture multiscale panoramic information—from tissue architecture down to individual organelles—providing structural information that can be used to build tissue-level digital cells with organelle-level detail. Nevertheless, due to the high-throughput and complex nature of EM, a robust computational framework for analyzing large-scale EM data is lacking. Some pioneering tools [21-27] have emerged to extract morphological and interaction features of organelles at nanoscale resolution. For example, mitoNet [21] and ERnet [22] enable accurate segmentation and reconstruction of specific types of organelles, such as mitochondria and endoplasmic reticulum (ER); built upon the Segment Anything Model (SAM), μ-SAM [28] provides a generalizable segmentation framework for both light and electron microscopy, achieving expert-level performance in mitochondrial segmentation; OpenOrganelle [24] enables whole cell-scale 3D organelle segmentation and MCSs analysis; and AIVE [26] enhances the quantification of organelle interactions through optimized boundary segmentation. In addition, to support high-throughput cross-sample contact analysis, we developed DeepContact, which simultaneously segments multiple organelles and statistically analyzes the functional relevance of MCSs [25]. However, these methods are limited in effective exploration of cell regions within large fields of view. Some recent studies intended to establish the relationship between organelles and their host cells by manually segmenting cell boundaries. For example, Jiang et al. [29] introduced a path-based method for automatically assigning mitochondria to individual cells, but it lacked the ability to distinguish between different cell types. Commonly used cell segmentation frameworks, such as Cellpose [30], Cellpose-SAM [31], Cellotype [32], and CellSAM [33], have not been specifically optimized for EM images. Furthermore, studies [34-37] have shown that organelle morphology and interactions vary significantly across different cell types and states, highlighting the importance of analyzing the spatial organization of organelles across cells.

To address these challenges and enable quantitative, high-throughput analysis of organelle interactions across different cell types at tissue and organ scales, we developed DeepOrganelle. It adopts a lightweight Mask2Former [38] framework as a universal segmentor and provides a highly efficient and generalizable framework for both 2D large-scale EM and 3D vEM segmentation and modeling. By performing cross-scale mapping, statistical quantification, panoramic characterization, and spatial visualization of intracellular organelles and their interactions across various cell types at the tissue level, DeepOrganelle creates a novel framework for digitizing the cell atlas at nanoscale resolution. This system provides an effective platform for the panoramic visualization and quantitative spatiotemporal profiling of organelles and their interactions within defined cellular populations in large-scale tissue EM datasets.

Germ cell differentiation and division are fundamental biological processes, yet systematic organelle-level analyses have been hindered by the absence of high-throughput, tissue-scale methodologies. Using DeepOrganelle, we performed high-throughput analysis and cross-scale visualization of organelles across an entire seminiferous epithelial cycle, encompassing 12 stages and 22 distinct statuses of germ cells. The results demonstrate that MCSs between the ER and mitochondria (ER-mito) and between the ER and plasma membrane (PM) undergo marked dynamic fluctuations across the seminiferous epithelial cycle. A subtle yet significant stepwise change in ER-mito MCSs was revealed within the pachytene, uncovering a previously unrecognized layer of subcellular heterogeneity within this critically important, yet conventionally homogenized, meiotic window. Noticeably, this finding is supported by a recent report that mitochondrial metabolic proteins, such as PDHA2, are essential for this transition [39]. Sertoli cells, as the sole somatic cell type in the seminiferous epithelium, maintain the immune-privileged environment and metabolic homeostasis essential for spermatogenesis. These cells exhibit exceptionally complex morphology and intricate ultrastructural features. Using DeepOrganelle, we delineated the blood–testis barrier (BTB) and multiple organellar networks of the intricate Sertoli cell networks *in situ*. This integrated approach enabled systematic quantification and visualization of spatiotemporal correlations between organelle redistribution and BTB remodeling dynamics. Thus, DeepOrganelle establishes an efficient and scalable analytical paradigm for high-throughput, cross-scale organelle profiling that is capable of discovering previously inaccessible nanoscale dynamics, such as dynamic organelle interactions and stage-specific reorganization. By bridging subcellular dynamics with tissue-level events, this tool creates new avenues for investigating ultrastructural networks in complex biological systems across diverse physiological and pathological contexts.

## Results

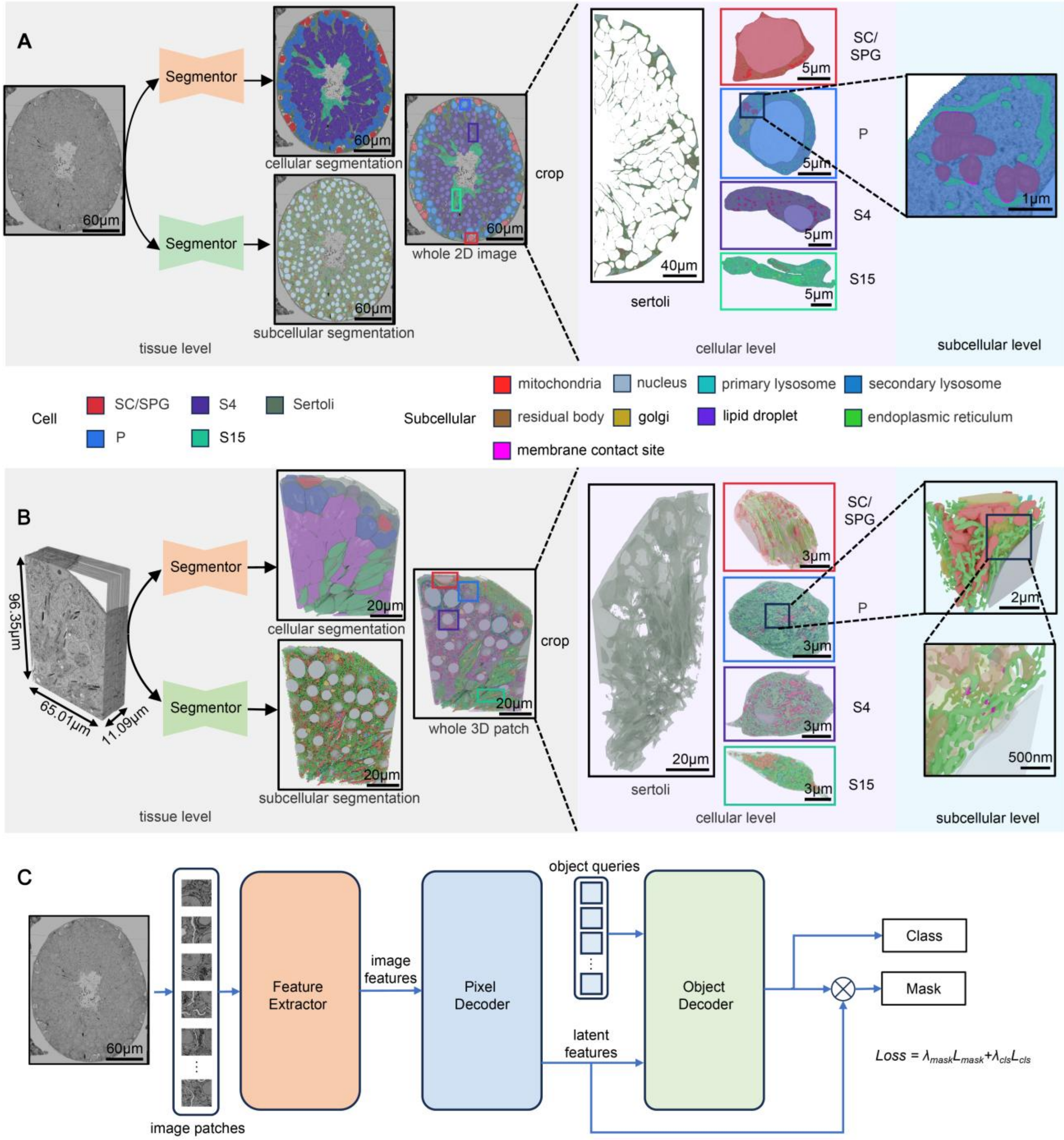

**Fig. 1: Workflow of DeepOrganelle.** (A, B) Multi-scale analysis of 2D large-scale electron microscopy (EM) (A) and 3D volume EM (B) data, showcasing views at the level of the whole 3D patch, single cells, and subcellular scale. SC/SPG, Stem cell/spermatogonia; P, pachytene spermatocytes; S4, step 4 spermatid; S16, step 16 spermatid. (C) The overall architecture of the panoptic segmentor in DeepOrganelle.

### *Workflow of DeepOrganelle*

DeepOrganelle was designed for the analysis of organelle distributions and quantification of contacts in 2D EM and 3D vEM at the tissue level. The system encompasses a uniform panoptic segmentation model as the segmentor and enables multi-level MCSs analysis for both 2D and 3D approaches.

The 2D EM workflow (**Fig. 1A**) encompasses trans-scale segmentation at both the cellular and organellar levels across an EM field of view spanning hundreds of micrometers. A unique ID was assigned to each segmented cell instance. Organelles and MCSs were quantified and recorded according to the corresponding cell instance ID. These IDs served as cell-level samples for analyzing organelle interactions, with all cells of the relevant type included to ensure comprehensive statistics.

The 3D EM analysis (**Fig. 1B**, Movie S1) was developed by extending the 2D quantification protocol with a layer-by-layer tracking algorithm to perform instance matching of cell and organelle instances. A fast-matching algorithm was designed using the Hungarian method to improve the efficiency of the 3D reconstruction process. Each 3D cell instance received a unique ID, and organelle and MCSs quantification was performed over the cell samples.

A lightweight panoptic segmentation model, Mask2Former [38], served as a universal segmentor for both cell and organelle segmentation (**Fig. 1C**). Employing a set-based prediction approach, it meticulously segments different types of subcellular organelles without boundary overlapping. To systematically profile the spermatogenic epithelial cycle, extensive annotations were built across 23 types of cells and 9 subcellular structures (Fig. S1). Comparative analysis (Table S1, S2) also showed that DeepOrganelle achieves state-of-the-art performance in segmenting both cellular and subcellular structures compared to other recent work.

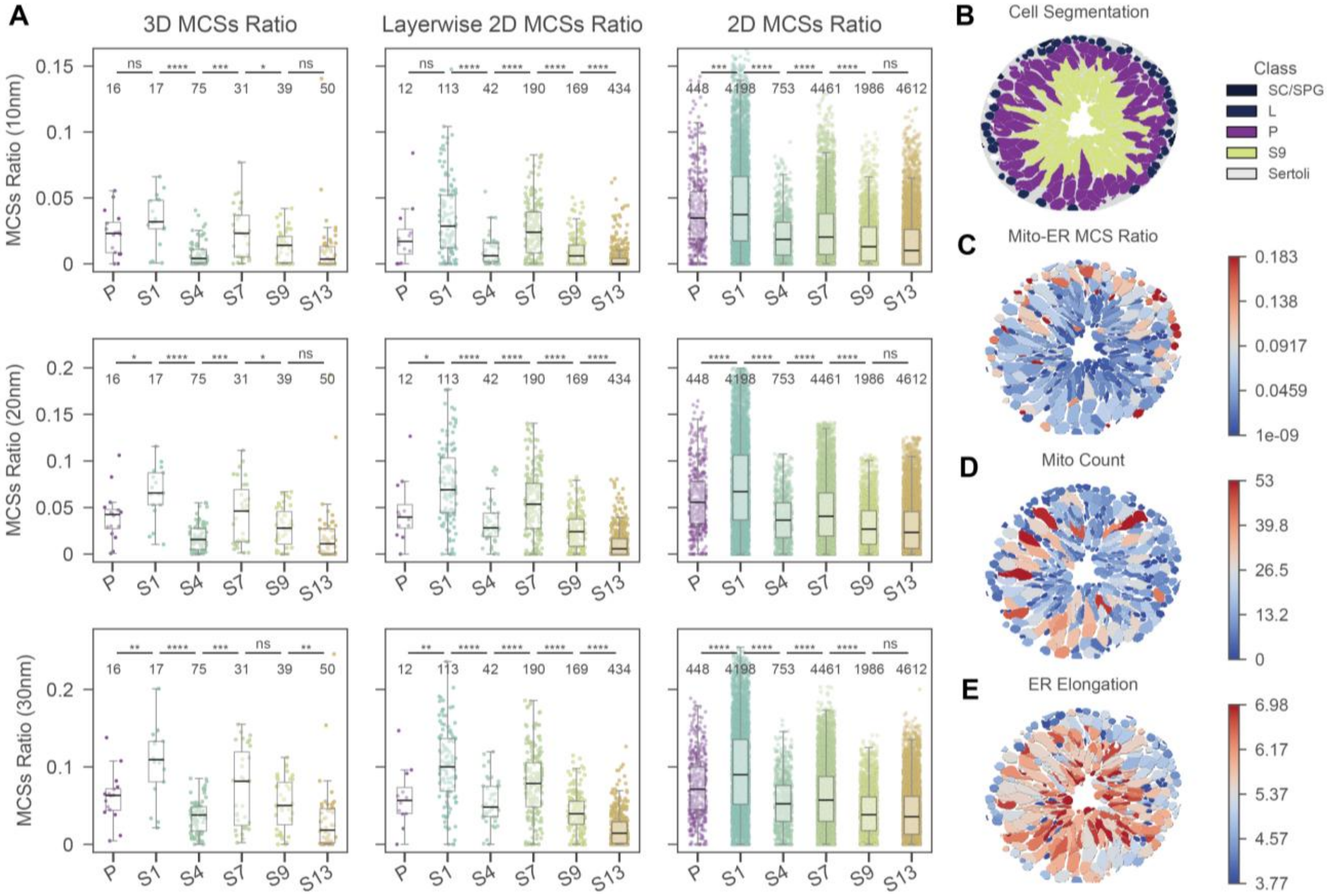

**Fig. 2: Cross-scale analysis of specific differentiation status of the germ cells in** the seminiferous epithelial tissue ***in situ*.** (A) ER-mito contact analysis in spermatogenic cells at different developmental stages, including pachytene spermatocytes (P) and step 1 (S1), step 4 (S4), step 7 (S7), step 9 (S9), step 13 (S13) spermatids. The levels of mitochondria-endoplasmic reticulume membrane contact sites (mito-ER MCSs) in various differentiation status were quantified using 3D contact analysis, layer-wise 2D contact analysis of volume electron microscopy (vEM) data, and 2D contact analysis of large-scale EM data separately, with mito-ER MCSs width cutoff at 10 nm, 20 nm, and 30 nm. (B) Spatial pattern of male germ cell differentiation of stage IX seminiferous epithelial tissue *in situ*. (C-E) Heat map quantification of mito-ER MCSs (C), Number of Mito (D) and ER elongation (E) in each single cell in stage IX seminiferous epithelial tissue.

### ***Cross-scale Analysis of Specific Cell Types***

Mouse testes have been studied intensively by EM, which has demonstrated a cycling pattern of the cellular and subcellular organization along with seminiferous epithelial cycling. The development and differentiation of spermatogenic cells in the mouse testis are precisely regulated. Typically, there are more than four differentiation statuses in male germ cell development on a cross-section of the seminiferous epithelium, and each step of differentiation is usually highly synchronized on cross-views. To explore the dynamic changes in ER-mito MCSs during male germ cell differentiation in spermatogenesis, comparative analyses between 2D and 3D EM images were conducted to ensure robust conclusions: 3D contact analysis of vEM, layer-wise contact analysis of vEM, and 2D large-scale contact analysis of EM. For 3D contact analysis, vEM data from spermatids at the pachytene spermatocyte, step1, step 4, step 7, step 9, and step 13 stages were analyzed to quantify the relative area of MCSs within each cell. This captured volumetric changes and revealed the 3D structural dynamics of MCSs for individual spermatids. In layer-wise contact analysis, vEM data were examined layer-by-layer with the contact ratio of MCSs computed and summed at each layer within individual cells, which enabled a direct comparison with the results from the 2D large-scale analysis and linked the 3D analysis to the 2D large-scale analysis. In 2D large-scale contact analysis, automated analysis was applied to large-scale EM images of seminiferous epithelium, capturing MCSs at pachytene spermatocyte, step1, step 4, step 7, step 9, and step 13. This focused on global spatial patterns and distribution trends across tissue sections. Importantly, the layer-wise contact analysis served as a bridge to validate the findings from the 2D analysis, ensuring consistency and enabling a more holistic understanding of MCSs dynamics.

Fig. 2A shows contact analyses performed using the three approaches, with cutoff distances set at 10 nm, 20 nm, and 30 nm. Results from both large-scale 2D and 3D analyses under all conditions consistently revealed a clear trend: as sperm cells advanced through developmental stages, particularly during spermatid maturation, the MCSs within the regressing cytoplasm decreased. This pattern, initially observed by 3D vEM, was corroborated by both layer-wise 2D vEM analysis and large-scale 2D EM analysis. The agreement across these methodological dimensions highlights the dynamic reorganization of MCSs as a key feature of spermatid differentiation and underscores the robustness of the experimental framework.

Fig. 2C-E shows a spatial heatmap and quantification analysis of different properties, including contact ratio (MCSs length/mitochondria perimeter) **(Fig. 2C)**, mitochondrial number **(Fig. 2D)**, and ER elongation ratio **(Fig. 2E)**. Notably, different cell types were arranged in a layered pattern along the basal-to-luminal axis of the seminiferous tubule, while properties within the same cell type remained relatively uniform.

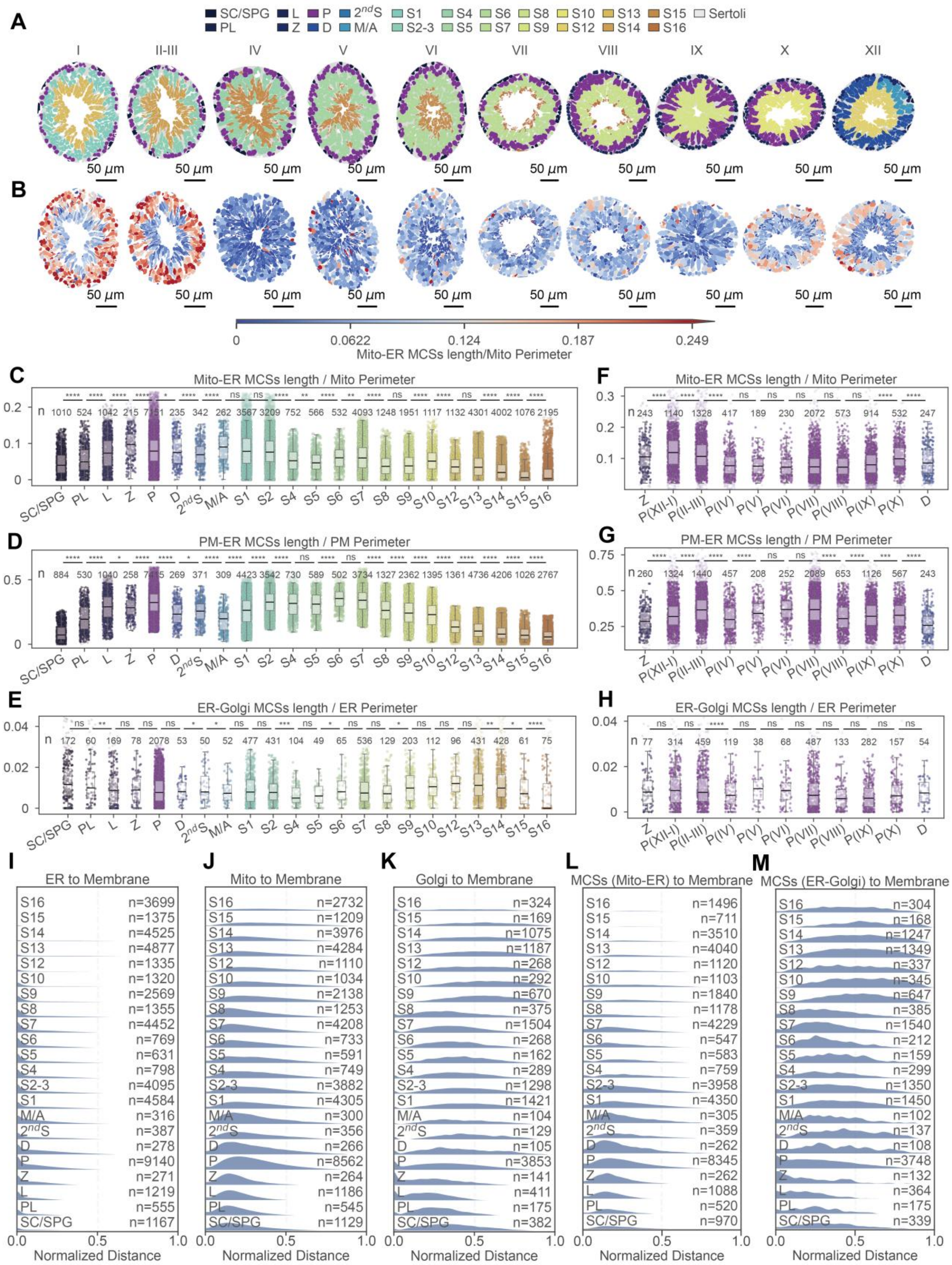

**Fig. 3 Systematic profiling of subcellular dynamics in spermatogenesis lineage.** (A) Differentiation atlas of the male germ cells during spermatogenesis. (B) Spatiotemporal heatmap

of ER-mito MCSs ratio for a whole spermatogenesis lineage *in situ*. (C-E) MCSs ratios for ER-mito (C), ER-plasma membrane (PM) (D), and Golgi-ER (E) during of the germ cells. (F-H) MCSs ratios for ER-mito (F), ER-PM (G), and Golgi-ER (H) within pachytene spermatocytes at different stages. (I-M) Distributions of ER (I), mitochondria (J), Golgi (K), Mito-ER MCSs (L) and ER-Golgi MCSs (L) with distance to membrane. SC/SPG, Stem cell/spermatogonia. PL, preleptotene spermatocytes; L, leptotene spermatocytes; Z, zygotene spermatocytes; P, pachytene spermatocytes; D, diplotene spermatocytes; $2^{nd}$S, secondary spermatocytes; M/A, Metaphase/anaphase; S1-S16, step 1-step 16 spermatid.

***Systematic Profiling subcellular dynamics during germ cell differentiation***

Meiosis and functional differentiation of the germ cells are fundamental questions in the life sciences. The seminiferous epithelium of the male gonad is a highly organized, cyclically patterned tissue in which germ cell differentiation follows strict spatiotemporal regulation [40]. The mouse seminiferous epithelium can be classified into 12 distinct stages (I–XII), each containing multiple synchronously differentiating phases of germ cells, including spermatogonia, spermatocytes in serial mitotic and meiotic status, and step-specific spermatids. The intricate architecture of the seminiferous epithelium necessitates a comprehensive approach to quantifying organelle interactions while accounting for spatial localization and diverse differentiation status.

Building on the comparative results of the 2D and 3D analyses in Fig. 2, we employed large-scale 2D EM to achieve panoramic profiling of the seminiferous epithelium along with the seminiferous cycling (Fig. 3a). Within the whole view of the seminiferous tissue in nanometer details with systemic EM imaging at various stages, DeepOrganelle discriminated 22 differentiation phases of germ cells in the process of spermatogenesis. As shown in **Fig. 3A**, germ cells were arranged in a layered pattern with stem cells/spermatogonia - primary/secondary spermatocytes - round/elongated spermatids along the basal-to-luminal axis of the seminiferous tubule. The layered patterning of germ cells transited gradually from one stage to the next, forming a spatiotemporal differentiation gradient of the germ cell differentiation atlas.

The contact ratio between the ER and mitochondria exhibited dynamic fluctuations throughout male germ cell differentiation. From spermatogonia to the zygotene stage, ER-mito MCSs levels progressively increased. A pronounced, unexpected decline was observed specifically during the pachytene stage. This was followed by a gradual decrease through diplotene and secondary spermatocytes, though a slight resurgence occurred during meiotic metaphase/anaphase. In the haploid spermatid phase (steps 1–16), ER-mito MCSs levels continued to display stage-specific variations: a steady decline from steps 1 to 5, an increase through step 6 to the top at step 7, an abrupt decrease at step 8, sustained low level at step 9, a minor recovery at step 10, and a final progressive decline until full sperm maturation at step 16 **(Fig. 3C).**

The sharp drop in ER-mito MCSs during pachytene was particularly notable, as this stage is characterized by active homologous recombination, checkpoint regulation, and synaptonemal complex disassembly—processes presumed to demand substantial metabolic and signaling coordination. To elucidate this paradox, we performed detailed pairwise comparisons between consecutive substages within pachytene. This analysis revealed a subtle yet defined waveform: ER-mito MCSs declined from pachytene stage I to its lowest level, which was sustained as a plateau from stages IV to VIII, before rising again at stages IX and X. Importantly, ER-mito MCSs levels in the preceding zygotene and subsequent diplotene stages were both significantly lower than those in adjacent pachytene substages (**Fig. 3F**).

ER-PM MCSs displayed finely regulated, wave-like fluctuations throughout spermatogenesis, with two prominent peaks at pachytene during meiosis and at step 6/7 in the spermatid differentiation process (**Fig. 3D**). Moreover, within pachytene phases (stages I–X), ER-PM MCS levels oscillated, reaching maxima at stages II–III and again at VI–VII, while showing a pronounced decline at stage IV (**Fig. 3G**). Together, these observations revealed a pattern of consecutive minor, yet significant, alterations in ER-PM MCSs, forming subtle waves that accompanied both meiotic progression and post-meiotic differentiation.

We further mapped the distribution of individual organelles and inter-organelle MCSs by measuring the distance of each object to the PM. Our analysis revealed a distinct cortical enrichment of the ER, most notably from step 1 to step 9 spermatids (Fig. 3I). In contrast, mitochondria were distributed more internally within the cytoplasm, showing only minor positional shifts between adjacent developmental stages. Notably, mitochondrial abundance peaked in prophase I (pachytene and diplotene) and decreased following meiotic divisions **(Fig. 3J**). ER-mito MCSs presented a similar spatial distribution as mitochondria but exhibited marked dynamics in amount; a sharp decline was observed in secondary spermatocytes at step 4/5, a transient increase at step 6/7, and a progressive and sustained decrease during the elongation phases of spermiogenesis beyond step 8 (**Fig. 3L**). The abundance of Golgi was stable throughout spermatogenesis until a sharp decrease with regression of the cytoplasm in the sperm maturation process (**Fig. 3M**). The positioning of the Golgi apparatus fluctuated between the plasma membrane and the nucleus across different stages (**Fig. 3K**), a pattern similarly observed in the spatial distribution of Golgi-ER membrane contact sites (MCSs) (Fig. 3M). Although the overall abundance of the Golgi varied somewhat among stages, the level of Golgi-ER MCSs remained relatively constant throughout spermatogenesis (**Fig. 3E, 3H, 3M**).

These nuanced patterns, revealed by the high sensitivity of DeepOrganelle detection, underscore the dynamic regulation of organelle interactions throughout meiotic prophase and highlight their potential role in supporting cellular differentiation at specific transitional points.

***Spatiotemporal Coordinations Between the dynamics of Sertoli Cell Organellar Network and the Blood-Testis Barrier Remodeling***

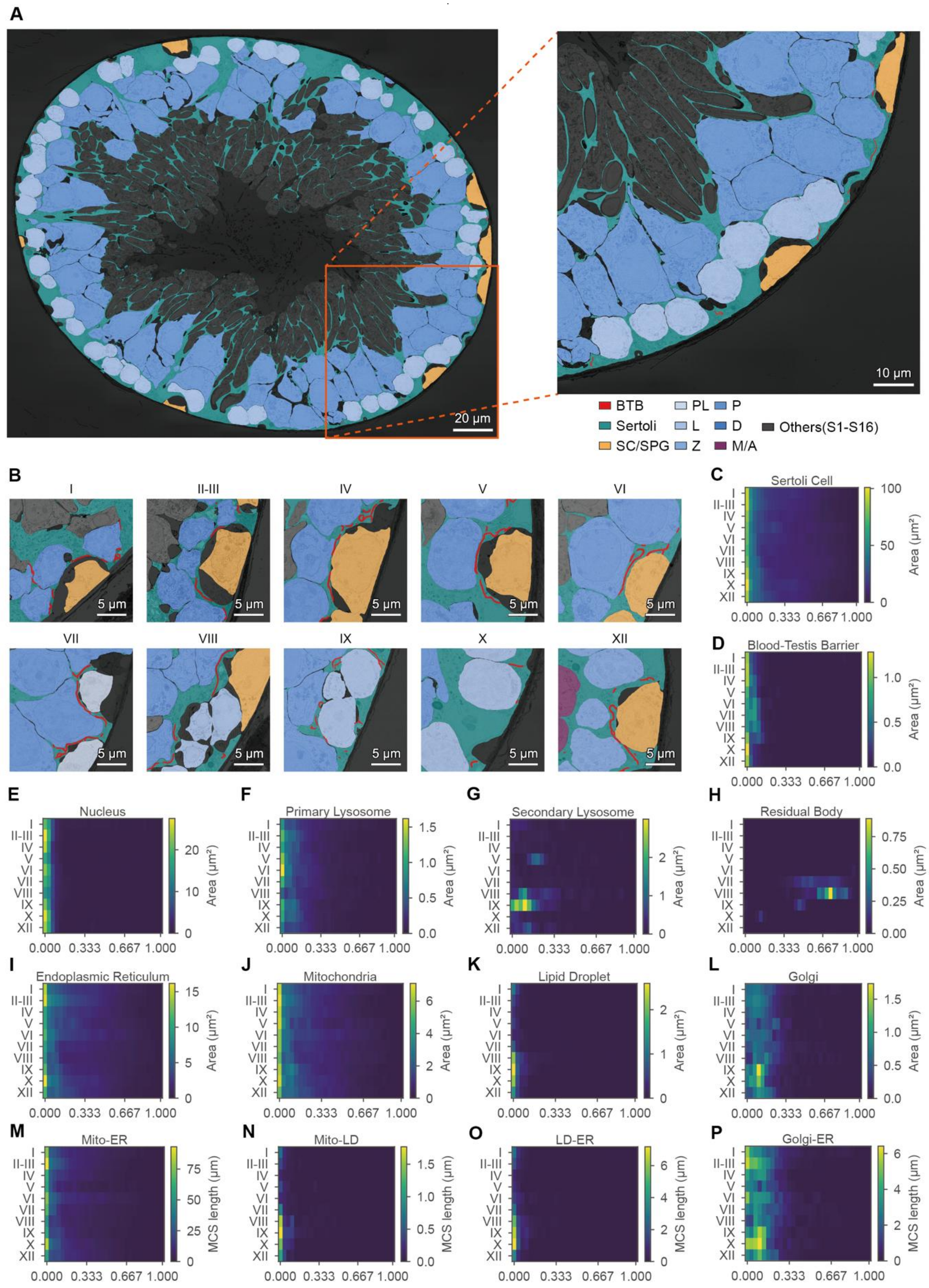

**Fig. 4: Spatiotemporal remodeling of Sertoli cell organelles**. (A) Visualization of the blood–testis barrier (BTB) in the whole seminiferous epithelial context. SC/SPG, Stem

cell/spermatogonia; PL, preleptotene spermatocytes; L, leptotene spermatocytes; Z, zygotene spermatocytes; P, pachytene spermatocytes; D, diplotene spermatocytes; $2^{nd}$S, secondary spermatocytes; M/A, Metaphase/anaphase; S1-S16, step 1-step 16 spermatid. (B) Enlarged view of the BTB and its basal seminiferous epithelium at different stages. The basal side is mostly facing to the right. (C) Radial distribution of Sertoli cell from the seminiferous epithelial edge to the lumen center. (D) Distributional probabilities of BTB from the seminiferous epithelial edge to the lumen center. (E-L) Distributional probabilities of different organelles from the seminiferous epithelial edge to the lumen center. (M-P) Distributional probabilities of different contacts from the seminiferous epithelial edge to the lumen center.

In the seminiferous epithelium, Sertoli cells play an essential regulatory role in the development and differentiation of germ cells [41]. By forming the BTB, they establish an immune-privileged environment for germ cells [42,43]. To elucidate the dynamic remodeling of the BTB, we trained a model using annotated key structural components—tight junctions and basal ectoplasmic specializations [41,44,45]. The BTB was predominantly positioned on the basal side of the epithelium. Together with the Sertoli cell body, this junctional complex effectively excluded stem cells and spermatogonia from the immune-protected compartment (**Fig. 4A, 4B**). Furthermore, preleptotene spermatocytes at stages VII and VIII were mainly situated outside the barrier (**Fig. 4B**). As these cells advanced to leptotene spermatocytes at stage IX, a new BTB assembled on the basal side of the leptotene spermatocyte prior to the disassembly of the old BTB on the adluminal side (**Fig. 4B**). By stage X, BTB restructuring was complete, and the original BTB on the adluminal side was fully resolved (**Fig. 4B, 4D**). This dynamic BTB remodeling during the preleptotene-to-leptotene transition in prophase I is consistent with its well-established role in creating and maintaining an immunologically protected microenvironment for meiotic and post-meiotic germ cell development (*37*).

The radial distribution of Sertoli cells (**Fig. 4C**) and their nuclei (**Fig. 4E**) demonstrated that the major cell bodies resided in the basal compartment of the epithelium with a minor shift between neighbor stages. Beyond this relatively consistent cellular distribution, detailed analysis of organelle distribution revealed pronounced stage-specific dynamics. Regarding lysosomal related structures, primary lysosomes (**Fig. 4F**) were most abundant at stage VI in the basal compartment, whereas secondary lysosomes (**Fig. 4G**) were predominantly enriched at stage IX, spatially corresponding to the location where the adluminal BTB disassembled. Residual bodies (**Fig. 4G)** were predominantly enriched at the adluminal side in stage VIII, spatiotemporally correlating with the releasing event of mature sperm at the adluminal side. The ER (**Fig. 4I**) was enriched specifically at stage I–III and stage X in basal Sertoli cells. Mitochondria (**Fig. 4J**) also showed a preferential distribution on the basal side. Notably, lipid droplets (LDs) (**Fig. 4K**) showed a pronounced distribution in basal Sertoli cells from stage VIII to stage X, with the peak at stage IX.

Corresponding to individual organelle distributions, the majority of ER-mito MCSs formed in the basal Sertoli cells, with significant stage-dependent fluctuations (**Fig. 4M**). Meanwhile, mito–LD and LD–ER MCSs (**Fig. 4N, 4O**) exhibited a predominant peak at stage IX and stage X, respectively, a pattern that aligns closely with the BTB-proximal redistribution of LDs (**Fig. 4K**). The amount of Golgi-ER MCSs (**Fig. 4P**) was also high at stage IX–X.

To summarize, our analysis revealed that, within the basal compartment of Sertoli cells, a coordinated, transient surge in the abundance of specific organelles, notably LDs, and their associated membrane contact sites occurs from stage VIII through stage X of spermatogenesis. This phase-specific reorganization coincides with the window of active BTB remodeling, underscoring a tightly synchronized, dynamic restructuring of the Sertoli cell organellar landscape in support of barrier function and tissue homeostasis.

## Discussion

DeepOrganelle employs a universal segmentor for both 2D large-scale EM and 3D vEM analysis. This unified approach enables a single model to continuously accumulate annotations of diverse cell types and organelles for various scenarios, eliminating the need for separate annotation efforts tailored to large-scale 2D or 3D EM datasets. This substantially enhances the model's generalizability and transferability. DeepOrganelle allows flexible selection of the 2D and/or 3D approach for digitizing organelles. The integration of 2D and 3D analyses in DeepOrganelle provides substantial advantages for large-scale EM studies. Using 2D imaging, researchers can cover broader scales, sample globally with higher throughput, and reduce instrument time and labeling costs compared to purely 3D approaches. This makes 2D analysis a cost-effective and efficient method for studying MCSs dynamics, especially in large-scale tissue-level investigations. The consistency between 2D and 3D analyses observed in this study (Fig. 2a) was also demonstrated in a previous study that evaluated the mitochondria-cristea ratio relative to mitochondria [46]. This further confirms the reliability of 2D methods as a proxy for 3D quantification, enabling scalable and high-resolution insights into organelle interactions within large-scale EM datasets. However, although 2D analysis yields ratio information comparable to 3D analysis, it is affected by cell polarity or tissue sample orientation during the analysis of organelle morphology. For large-scale tissue EM analysis, it is essential to control the image acquisition plane and axis at the same level among experimental groups, especially for the evaluation of non-normalized parameters, such as organelle morphology. We suggest adopting 3D quantification as a complement to 2D methods for some typical situations when the morphology and distributions of organelles are anisotropical.

DeepOrganelle was used to digitize the seminiferous epithelial cycle at nanoscale resolution, demonstrating its potential to revolutionize cell atlas and lineage studies by capturing subcellular dynamics from an integrated, whole-tissue perspective. By leveraging high-throughput 2D tissue EM sampling, our approach enabled the discovery of previously inaccessible, stage-resolved organelle dynamics. Specifically, we identified a stereotypic fluctuation of ER-mito contact sites in which they progressively declined through early pachytene, remained low at mid-stages, and peaked sharply at stage X—a spatiotemporal pattern strongly suggestive of active regulatory control. This fine, phase-specific mapping revealed that the pachytene stage, conventionally defined by chromatin morphology, harbors delicate, cyclical variations in organelle contact that correlate precisely with the seminiferous epithelial cycle. The contiguous and oscillatory nature of ER–mito contact across meiosis, faithfully captured by our large-scale analytical pipeline, underscores the power of this integrated framework to resolve dynamic subcellular architectures. Notable distinct peaks in both ER-mito and ER-PM MCSs at stage X pachytene followed by a sharp decline upon transition to diplotene were systematically quantified by our platform. These structural dynamics align with recent functional evidence that mitochondrial metabolic proteins, such as PDHA2, essential for ATP production during double-strand break repair, are required for this meiotic transition [39]. Thus, DeepOrganelle not only documents subcellular reorganization but provides a quantitative, scalable framework to link ultrastructural dynamics with molecular mechanisms *in vivo*.

Previous manual counting-based EM quantification studies have revealed dynamics of LDs during the seminiferous epithelial cycle in different species [47]. Such dynamics of the organellar networks likely play an important role in maintaining homeostasis of the seminiferous epithelial tissue [48]. Although DeepOrganelle confirmed dynamic restructuring of the BTB during the preleptotene-to-leptotene transition, consistent with its role in maintaining an immuno-privileged environment (Fig. 4B), it also revealed an unexpected spatial reorganization: LDs and their contacts with ER and mitochondria exhibited a distinct redistribution peak precisely during BTB remodeling. These findings demonstrate a unique, stage-specific reorganization of organelle topography in support of BTB dynamics. Notably, lysosome-mediated residual body recycling occurs just prior to BTB restructuring and involves a transition from apical to basal localization (Fig. 4G, 4H). This process may provide a critical lipid source to support metabolism in the basal compartment. Spatiotemporal analysis indicated that such synchronized coordination of organellar networks likely plays a key role in maintaining homeostasis of the seminiferous epithelium. The integration of such ultrastructural functional insights with emerging multi-omics approaches, such as single-cell spatial transcriptomics [49,50] and metabolomics [51,52], will generate an unprecedented, precise functional atlas of spermatogenesis.

Looking ahead, DeepOrganelle meets the growing demand for large-scale EM and vEM in biomedical research and enables high-throughput cross-sample quantitative functional analysis of organelles and MCSs. This tool also opens an avenue for integrating ultrastructural networks into large-scale physiological and pathological contexts. Such ultrastructural information can be further integrated with single-cell transcriptomics, proteomics, and other omics to construct digital cells. This integration is poised to advance the construction of a "queryable" and "predictive" digital cell system with spatial-physical constraints and realistic subcellular dynamics. In sum, DeepOrganelle stands as a promising routine tool to drive digital cell construction and cell atlas and lineage studies into a subcellular level.

**Acknowledgements**

We would like to thank Zihan Chen from Beijing University of Posts and Telecommunications (BUPT) for algorithm testing. Shiyu Zhan from BUPT for system development. Xueke Tan, Tianjiao Wang, Yun Feng from the Center for Biological Imaging, Institute of Biophysics (IBP), Chinese Academy of Sciences (CAS) for their help with sample preparation and large-scale EM imaging and valuable discussions on 3D visualization. Yunqi Wang, Xiaopeng Li from IBP for labeling. Wenjing Li from Institute of Automation, CAS for valuable discussions on organelle biology.

**Funding**

This study was funded by the National Natural Science Foundation of China(92354307, 32525005, 92254306), National Key Research and Development Project of China(2021YFA1301500), and the Fundamental Research Funds for the Beijing University of Posts and Telecommunications under Grant 2025AI4S19.

**Author Contributions**

Conceptualization: LX, LQL, GY, and TX; Methodology:LX, HJW; Data Collections and Annotations:LQL, JYZ, HJW,XXL; Software: HJW and LX; Investigation: LX,LQL and HJW; Visualization: HJW, LQL, LX, and JJH; Funding acquisition: GY, TX, FS and LX; Project administration: TX, GY, FS, and LX; Supervision: LX, TX, GY, FS, JJH, LQL, and YZ; Writing – original draft: LX, HJW, and LQL; Writing – review & editing: LX, LQL, GY and HJW.

**Competing Interests**

The authors declare no competing of interests.